\documentclass[12pt,letterpaper,english]{article}
\usepackage[T1]{fontenc}
\usepackage{babel}
\usepackage{graphics}

\makeatletter

\providecommand{\LyX}{L\kern-.1667em\lower.25em\hbox{Y}\kern-.125emX\@}

\usepackage[T1]{fontenc}


\makeatother
\begin{document}
\baselineskip .3in 

\begin{titlepage}

{\centering \textbf{\large Dynamic critical behavior of failure and
plastic deformation in the random fiber bundle model }\large \par}

\vskip .1in 

{\noindent \centering \textbf{S. Pradhan} \( ^{(1)} \)\textbf{, P.
Bhattacharyya}\( ^{(2)} \) and \textbf{B. K. Chakrabarti} \( ^{(3)} \)\par}

{\noindent \centering \textit{Saha Institute of Nuclear Physics},\\
 \textit{1/AF Bidhan Nagar, Kolkata700 064, India.}\\
 \par}

\noindent \vskip .1in

\noindent \textbf{Abstract}

\noindent The random fiber bundle (RFB) model, with the strength of
the fibers distributed uniformly within a finite interval, is studied
under the assumption of global load sharing among all unbroken fibers
of the bundle. At any fixed value of the applied stress \( \sigma  \)
(load per fiber initially present in the bundle), the fraction \( U_{t}(\sigma ) \)
of fibers that remain unbroken at successive time steps \( t \) is
shown to follow simple recurrence relations. The model is found to
have stable fixed point \( U^{\star }(\sigma ) \) for applied stress
in the range \( 0\leq \sigma \leq \sigma _{c} \), beyond which total
failure of the bundle takes place discontinuously (abruptly from \( U^{\star }(\sigma _{c}) \)
to \( 0 \)) . The dynamic critical behavior near this \( \sigma _{c} \)
has been studied for this model analysing the recurrence relations.
We also investigated the finite size scaling behavior near \( \sigma _{c} \).
At the critical point \( \sigma =\sigma _{c} \), one finds strict
power law decay (with time \( t \)) of the fraction of unbroken fibers
\( U_{t}(\sigma _{c}) \) (as \( t\rightarrow \infty  \)). The avalanche
size distribution for this mean-field dynamics of failure at \( \sigma <\sigma _{c} \)
has been studied. The elastic response of the RFB model has also been
studied analytically for a specific probability distribution of fiber
strengths, where the bundle shows plastic behavior before complete
failure, following an initial linear response. \\
 \vskip .1in

\noindent \textit{e-mail addresses} :

\noindent \( ^{(1)} \)spradhan@cmp.saha.ernet.in

\noindent \( ^{(2)} \)pratip@cmp.saha.ernet.in

\noindent \( ^{(3)} \)bikas@cmp.saha.ernet.in 

\noindent \end{titlepage}

\noindent \textbf{I. Introduction }

\noindent A typical relaxational dynamics has been observed in a strained
random fiber bundle (RFB) model {[}1-8{]} where \( N \) fibers are
connected in parallel to each other and clamped at their two ends
and the strength of the individual fibers has some particular distribution
(white, Gaussian or otherwise). In the global load-sharing approximation
\cite{Dan45,HH92}, at any instant, the surviving fibers all share
equally the external load (irrespective of their proximity etc. of
the fiber to failed fibers etc.). Initially, after the load \( F \)
is applied on the bundle, a fraction of the fibers having strength
less than the applied stress \( \sigma =F/N \) fail immediately.
After this, the total load on the bundle redistributes globally as
the stress is transferred from broken fibers to the remaining unbroken
ones. This redistribution causes secondary failures which in general
causes further failures and so on. After some relaxation time \( \tau  \),
which depends on \( \sigma  \), the system ultimately becomes stable
if the applied stress \( \sigma  \) is less than or equal to a critical
value \( \sigma _{c} \), and beyond which (\( \sigma  \) \( > \)
\( \sigma _{c} \)) all the fibers break and the bundle fails completely.
Although the local load sharing might be more realistic, we study
here the global load sharing model because of its simplicity. The
study of the scaling properties of the dynamics of the fiber bundle
model systems is expected to be extremely useful in analysing the
statistics of fracture and breakdown in real materials, including
earthquakes \cite{BB97,Bak97}.

In this paper, we report on the critical dynamics of the RFB model
in the global load-sharing case, assuming uniform distribution of
threshold strength of the fibers (up to a cutoff), in particular at
the critical point \( \sigma _{c} \). In a previous paper \cite{SB01},
we have solved the dynamics of the model, showing a novel critical
behavior as the stress \( \sigma  \) approaches a certain value \( \sigma _{c} \);
we had derived there the expressions for the breakdown susceptibility
\( \chi  \) and the relaxation time \( \tau  \) under a stress \( \sigma  \)
\( <\sigma _{c} \) and showed that both the quantities diverge following
power laws as \( \sigma  \) approaches \( \sigma _{c} \) from below.
Here we define an order parameter for the transition from a state
of partial failure of the bundle to a state of total failure and also
show that at the critical stress \( \sigma _{c} \), the dynamics
follows a precise and strict power law. From the finite size dependence
of \( \sigma _{c} \) and the order parameter we have identified the
correlation length exponent of the system. We have studied the avalanche
size statistics in the model as well. Considering a modified (uniform
but shifted from the origin) distribution of fiber strengths we have
studied analytically the elastic-plastic deformation characteristics
\cite{More94} of the RFB model. 

\vskip .1in

\noindent \textbf{II. The model}

\noindent The RFB model consists of \( N \) elastic fibers clamped
at two ends (Fig. 1), where the failure stress of the individual fibers
are distributed randomly and uniformly in the interval between \( 0 \)
and \( 1 \) (white or uniform distribution; Fig. 2). Global load
sharing is assumed; i.e., the applied load on the bundle is equally
shared among all the existing intact fibers. This assumption neglects
`local' fluctuations in stress (and its redistribution) and renders
the 

\vspace{0.3cm}
{\centering \resizebox*{9cm}{5cm}{\rotatebox{270}{\includegraphics{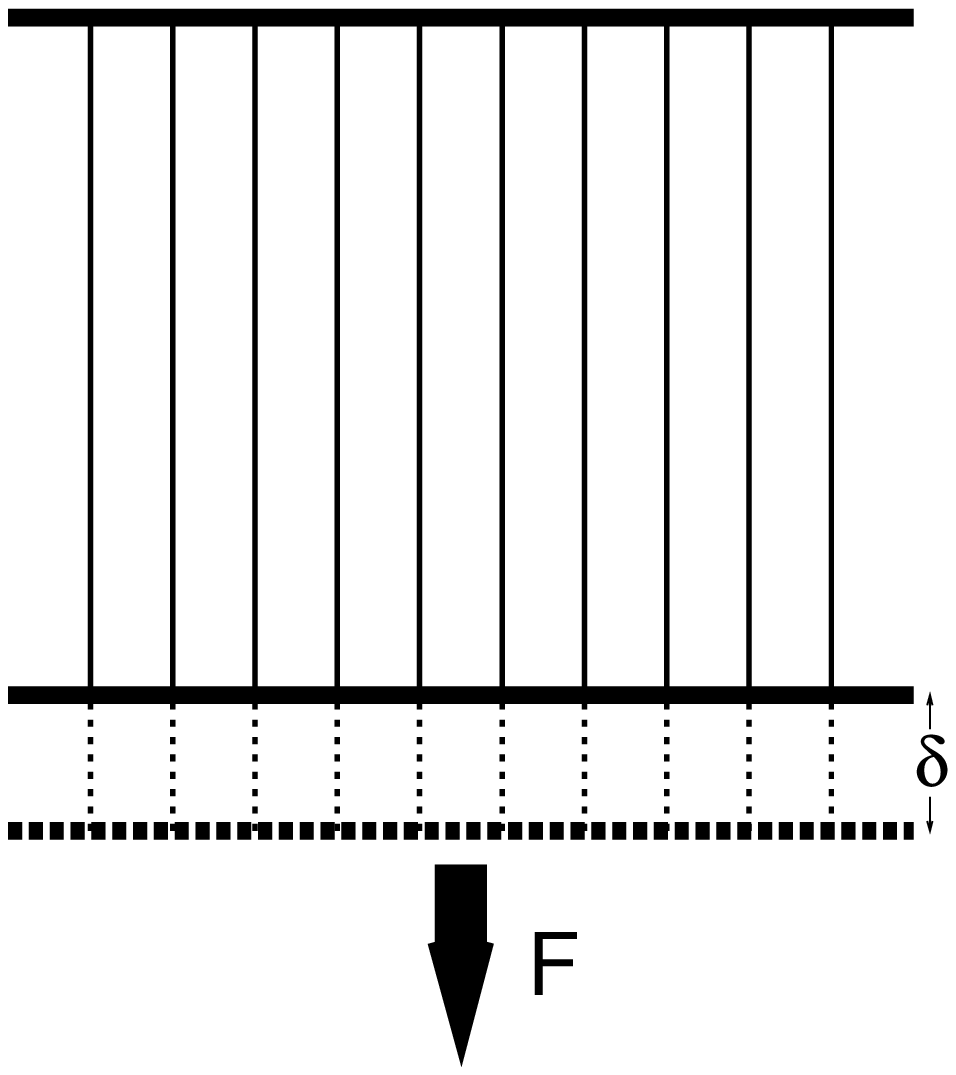}}} \par}
\vspace{0.3cm}

\noindent \textbf{\footnotesize Fig. 1:} {\footnotesize The RFB model
consists of \( N \) fibers. The bundle is subjected to a load \( F \).
Assuming linear elasticity, with identical elastic constant \( \kappa  \)
for each of the fiber up to the breaking, the load \( F \) can be
expressed as \( N\kappa \delta  \) where \( \delta  \) denotes the
strain for the fibers until any of them breaks. The breaking strengths
of the fibers are assumed to be random, as discussed later.}{\footnotesize \par}

\vskip .1in

\vspace{0.3cm}
{\centering \resizebox*{8cm}{5cm}{\includegraphics{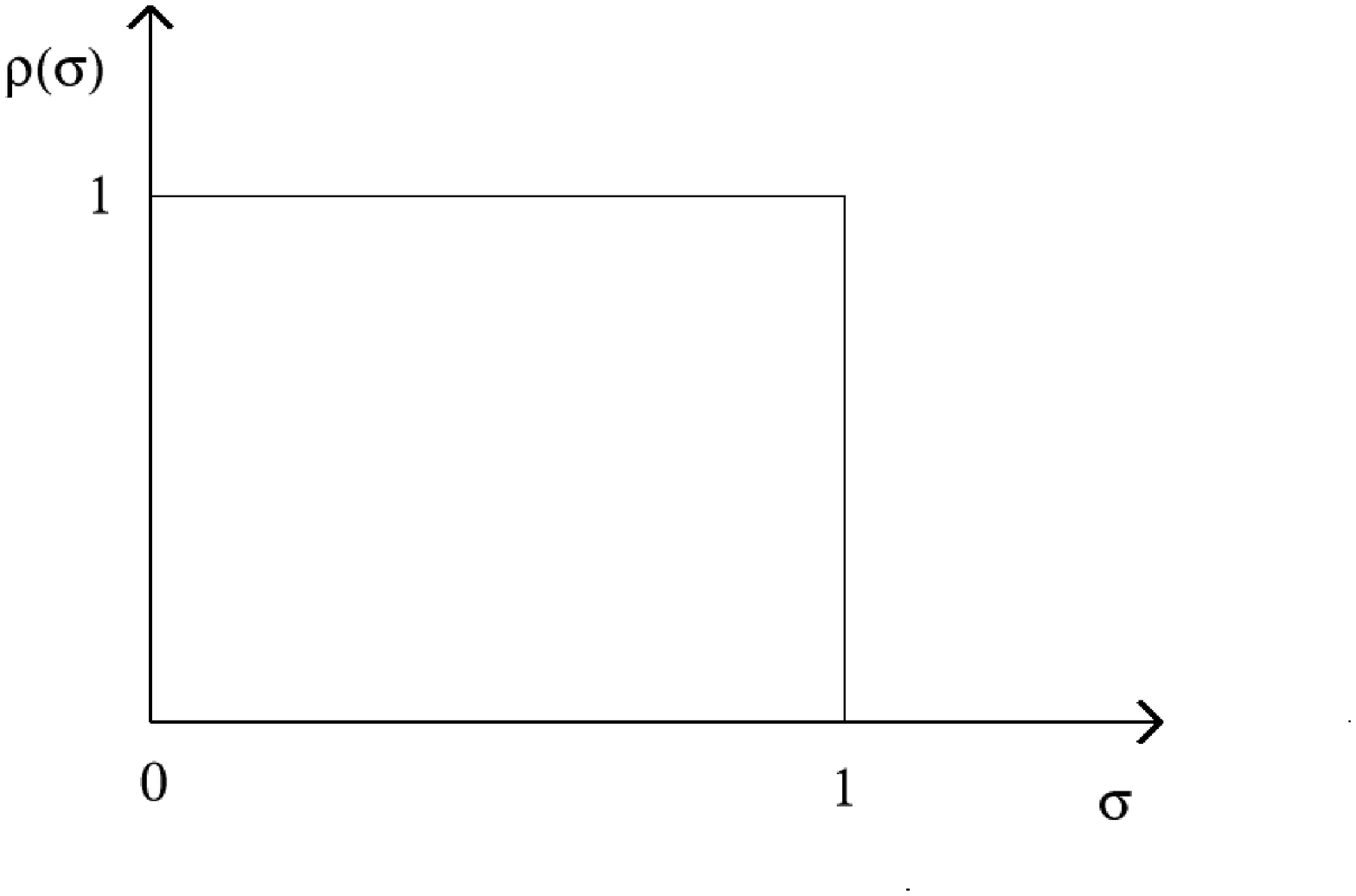}} \par}
\vspace{0.3cm}

\noindent \textbf{\footnotesize Fig. 2:} {\footnotesize The simplest
model considered here assumes uniformly random distribution or white
distribution \( \rho (\sigma ) \) for the strength of the fibers
up to a (normalised) cutoff strength. This distribution gives the
recurrence relation (1).}{\footnotesize \par}

\vskip .1in

\noindent model as a mean-field one. We work with the fraction \( U_{t} \)
\( \equiv N_{t}/N \); \( N_{t} \) being the number of fibers remaining
intact after \( t \) time-steps and \( N_{t=0}=N \). 

With the application of any small load \( F \) (\( = \) \( \sigma  \)
\( N \), with \( \sigma \ll 1 \)) on the bundle, an initial stress
\( \sigma  \) (load per fiber) sets in. At this first step therefore,
\( \sigma  \)\( N \) number of fibers break, leaving \( NU_{1}( \)\( \sigma ) \)
\( = \) \( N(1-\sigma ) \) unbroken fibers. After this, the applied
force is redistributed uniformly among the remaining intact fibers
and the redistributed stress becomes \( F/ \)\( [NU_{1}( \)\( \sigma )] \)
\( = \) \( \sigma / \)\( (1-\sigma ) \). Some more fibers, for
which the strengths are below the value of the redistributed stress,
fail thus leaving \( NU_{2}( \)\( \sigma ) \) \( = \) \( N[1-\sigma /(1-\sigma )] \)
unbroken fibers. This in turn increases the redistributed stress and
induces further failures. Consequently, as the stress per fiber \( \sigma (t) \)
at time \( t \) is given by \( F/NU_{t}=\sigma /U_{t} \) and the
surviving fraction is given by \( 1-\sigma /U_{t} \) (see Fig. 2),
\( U_{t}(\sigma ) \) follows a simple recurrence relation \begin{equation}
\label{qrw}
U_{t+1}(\sigma )=1-\frac{\sigma }{U_{t}(\sigma )}.
\end{equation}

\vskip.1in

\noindent \textbf{III. Breaking dynamics of the RFB model }

\noindent The recurrence relation (1) has the form of an iterative
map \( U_{t+1}=Y(U_{t}) \). Its fixed point \( U^{\star } \) is
defined by the relation: \( U^{\star }=Y(U^{\star }) \) and from
eqn. (1) one gets

\noindent \begin{equation}
\label{qq}
U^{\star }(\sigma )=\frac{1}{2}\pm (\sigma _{c}-\sigma )^{1/2};\sigma _{c}=\frac{1}{4}.
\end{equation}

\noindent The quantity \( U^{\star } \) must be real valued as it
has a physical meaning: it is the fraction of the original bundle
that remains intact under a fixed applied stress \( \sigma  \) when
the applied stress lies in the range \( 0\leq \sigma \leq \sigma _{c} \).
For \( \sigma >\sigma _{c} \) the map does not have a real-valued
fixed point and as can be seen from (1), the dynamics never stops
until \( U_{t}=0 \) when the bundle breaks completely. Since it requires
that \( |dY/dU|_{U^{\star }(\sigma )} \) <1 for a fixed point \( U^{\star }(\sigma ) \)
to be stable, for each value of \( \sigma  \) the value of \( U^{\star } \)
with the positive sign in eqn. (2) represents a stable fixed point
(or attractor) while the value of \( U^{\star } \) with the negative
sign in eqn. (2) represents an unstable fixed point (or repeller).
It may be noted that the quantity \( U^{\star }-1/2 \) behaves like
an order parameter that determines a transition from a state of partial
failure (\( \sigma \leq \sigma _{c} \)) to a state of total failure
(\( \sigma >\sigma _{c} \)) : \begin{equation}
\label{dec31}
O\equiv U^{\star }-1/2=(\sigma _{c}-\sigma )^{\beta };\beta =\frac{1}{2}.
\end{equation}

\newpage

\noindent \textbf{IV. Critical behavior}

\noindent \textbf{(a) For \( \sigma <\sigma _{c} \)}

\noindent To study the dynamics away from criticality (\( \sigma \rightarrow \sigma _{c} \)
from below), we replace the recurrence relation (1) by a differential
equation \begin{equation}
\label{qwes}
-\frac{dU}{dt}=\frac{U^{2}-U+\sigma }{U}.
\end{equation}

\noindent Close to the fixed point we write \( U_{t}(\sigma )=U^{\star }(\sigma ) \)
+\( \epsilon  \) which, following eqn. (4), gives \cite{SB01} \begin{equation}
\label{qas}
\epsilon =U_{t}(\sigma )-U^{\star }(\sigma )\approx \exp (-t/\tau ),
\end{equation}

\noindent where \( \tau =\frac{1}{2}\left[ \frac{1}{2}(\sigma _{c}-\sigma )^{-1/2}+1\right]  \).
Approaching \( \sigma _{c} \) from below we get \begin{equation}
\label{dec19}
\tau \propto (\sigma _{c}-\sigma )^{-\alpha };\alpha =\frac{1}{2}
\end{equation}
 as the relaxation time of the model and it is found to diverge following
a power-law as \( \sigma \rightarrow \sigma _{c} \) from below. Although,
we have used here the continuum-time version (4) of the recurrence
relation to evaluate the relaxation time (\( \tau  \)), we have checked
numerically as well from the discrete-time recurrence relation (1)
and obtained the same exponent value. 

One can also consider the breakdown susceptibility \( \chi  \), defined
as the number (fraction) of fibers that break due to an infinitesimal
increment of the applied stress \cite{SB01} \begin{equation}
\label{sawq}
\chi =\left| \frac{dU^{\star }(\sigma )}{d\sigma }\right| =\frac{1}{2}(\sigma _{c}-\sigma )^{-\gamma };\gamma =\frac{1}{2}
\end{equation}

\noindent from equation (2). Hence \( \chi  \) too diverges as the
applied stress \( \sigma  \) approaches the critical value \( \sigma _{c}=\frac{1}{4} \).
Such a divergence in \( \chi  \) had already been observed in the
numerical measurements \cite{PS97,RS99}. 

\noindent \textbf{(b) At \( \sigma =\sigma _{c} \)}

\noindent At \( \sigma =\sigma _{c} \) the fraction of fibers surviving
is \( U^{\star }(\sigma _{c})=\frac{1}{2} \) and \( |dY/dU|_{U^{\star }(\sigma _{c})}=1 \)
which suggests that the system will take infinite time to reach the
fixed point at \( \sigma _{c} \). From the recurrence relation (1)
it can be shown that this decay of the fraction \( U_{t}(\sigma _{c}) \)
of unbroken fibers that remain intact at time \( t \) follows a simple
power-law: 

\begin{equation}
\label{qqq}
U_{t}=\frac{1}{2}(1+\frac{1}{t+1}),
\end{equation}

\noindent starting from \( U_{0}=1 \). For large \( t \) (\( t\rightarrow \infty  \)),
this reduces to \( U_{t}-1/2\propto t^{-1} \); a simple but strict
power law. 

\noindent \vskip .1in

\noindent \textbf{V. Finite size effects and correlation length exponent}

\noindent For a finite bundle of \( N \) fibers, the recurrence relation
(1) will be replaced by \begin{equation}
\label{dec31}
N_{t+1}(\sigma )=N-\left\lfloor \frac{N^{2}\sigma }{N_{t}}\right\rfloor ,
\end{equation}
 where \( \left\lfloor x\right\rfloor  \) denotes the greatest integer
less than or equal to \( x \). Here the fixed point is obtained when
\( N_{t+1}=N_{t}=N^{\star } \) and the value of \( N^{\star } \)
is bounded by the relation \begin{equation}
\label{dec31}
N\left[ \frac{1}{2}+\left( \frac{1}{4}-\sigma \right) ^{1/2}\right] \leq N^{\star }<\frac{1}{2}\left( N+1\right) +\left[ \frac{1}{4}(N+1)^{2}-N^{2}\sigma \right] ^{1/2},
\end{equation}
 which clearly depends on the finite size of the system. Consequently
the effective critical point \( \sigma _{c}(N) \) for the finite
RFB model is bounded as: \begin{equation}
\label{dec28}
\frac{1}{4}\leq \sigma _{c}(N)<\frac{1}{4}[1+\frac{1}{N}]^{2}.
\end{equation}
 It follows from eqn. (10) that, at \( \sigma _{c}=\frac{1}{4} \),
the fixed point value \( N^{\star } \) for a finite bundle decays
with the initial bundle size \( N \) following a power law \begin{equation}
\label{dec31}
N_{\sigma _{c}=1/4}^{\star }-\frac{N}{2}\sim N^{1/2}.
\end{equation}

\noindent Since the quantity \( U^{\star }-1/2 \) in eqn. (3) behaves
like an order parameter for a phase transition, the corresponding
quantity in a finite bundle of \( N \) fibers would be \begin{equation}
\label{dec31}
\frac{N^{\star }-\frac{N}{2}}{N}\equiv U^{\star }_{N}-\frac{1}{2}.
\end{equation}

\noindent Expressing the correlation length as \( \xi \propto (\sigma _{c}-\sigma )^{-\nu } \)
in the infinite system and combining it with eqn. (3) for a finite
size system (where \( \xi \sim N \)), the finite size scaling behavior
can be written as \begin{equation}
\label{dec31}
U_{N}^{\star }(\sigma _{c})-\frac{1}{2}\sim N^{-\beta /\nu }.
\end{equation}

\noindent Since \( \beta =1/2 \), as obtained earlier from eqn. (3),
we get \( \nu =1 \) by comparing eqn. (14) with eqn. (12). 

\newpage

\noindent \textbf{VI. Avalanche size distribution}

\noindent We now study the avalanche size distribution in this mean-field
model. If one considers strictly uniform strength distribution of
the fibers in this model, one can not meaningfully approach the failure
point by breaking the weakest fiber and looking for the avalanches
of successive failures of the fibers, following the avalanche definition
of Hemmer et al \cite{HH92,D92}. If we apply this definition in the
above (restricted) model, we will end up with only two distinct sizes
of avalanches: \( (N/2) \) avalanches of unit size and one avalanche
of size \( (N/2) \). This will occur due to the perfectly uniform
strength distribution of the fibers (with the successive strength
of fibers differing by \( 1/N \)). To work therefore with a more
general definition of avalanche , we increase the external load on
the bundle steadily such that the external load \( F \) increases
by an equal amount (\( dF=Nd\sigma  \)) at each step (cf. \cite{Pach2000}).
This ensures the bimodal, yet decreasing, distribution function mentioned
above to become a smooth (decaying) function. Operationally also,
this procedure is quite common and can be applied to different cases
and to bundles with different types of strength distribution \( \rho (\sigma ) \)
of fibers. Here, the fraction of fibers \( m \) which eventually
fail due to this increase in load or stress may be considered as the
avalanche size: \begin{equation}
\label{feb5}
m=\frac{dM}{d\sigma };M=1-U^{\star }(\sigma ).
\end{equation}

\noindent With \( U^{\star }(\sigma ) \) from (2) we get \begin{equation}
\label{feb5}
\sigma _{c}-\sigma \sim m^{-2}.
\end{equation}

\noindent If we now define the avalanche size probability distribution
by \( P(m) \), then \( P(m)\Delta m \) measures \( \Delta \sigma  \),
the number of times one has to change \( \sigma  \) (by \( d\sigma  \))
to get a change \( \Delta m \) along the \( m \) versus \( \sigma  \)
curve in (16). In other words, \begin{equation}
\label{feb5}
P(m)=\frac{d\sigma }{dm}\sim m^{-\eta };\eta =3.
\end{equation}

This mean-field result for \( P(m) \) (power law decay with exponent
\( \eta =3 \)) is obtained here for global load sharing and uniform
fiber strength distribution when the external load is increased by
a fixed amount. We have checked this result numerically for different
\( d\sigma  \) values ( \( =1/N \)) for bundles with \( 50,000 \)
fibers having both strictly uniform and uniform-on-average strength
distributions. The results are shown in Fig. 3. The earlier numerical
results of Moreno et. al. \cite{Pach2000} for Weibull type distribution
of fiber strength also confirms the relation (16), which implies that
the cumulative distribution decreases with avalanche size \( m \)
as \( m^{-2} \), in agreement with (17).

\vspace{0.3cm}
{\centering \resizebox*{12cm}{10cm}{\rotatebox{-90}{\includegraphics{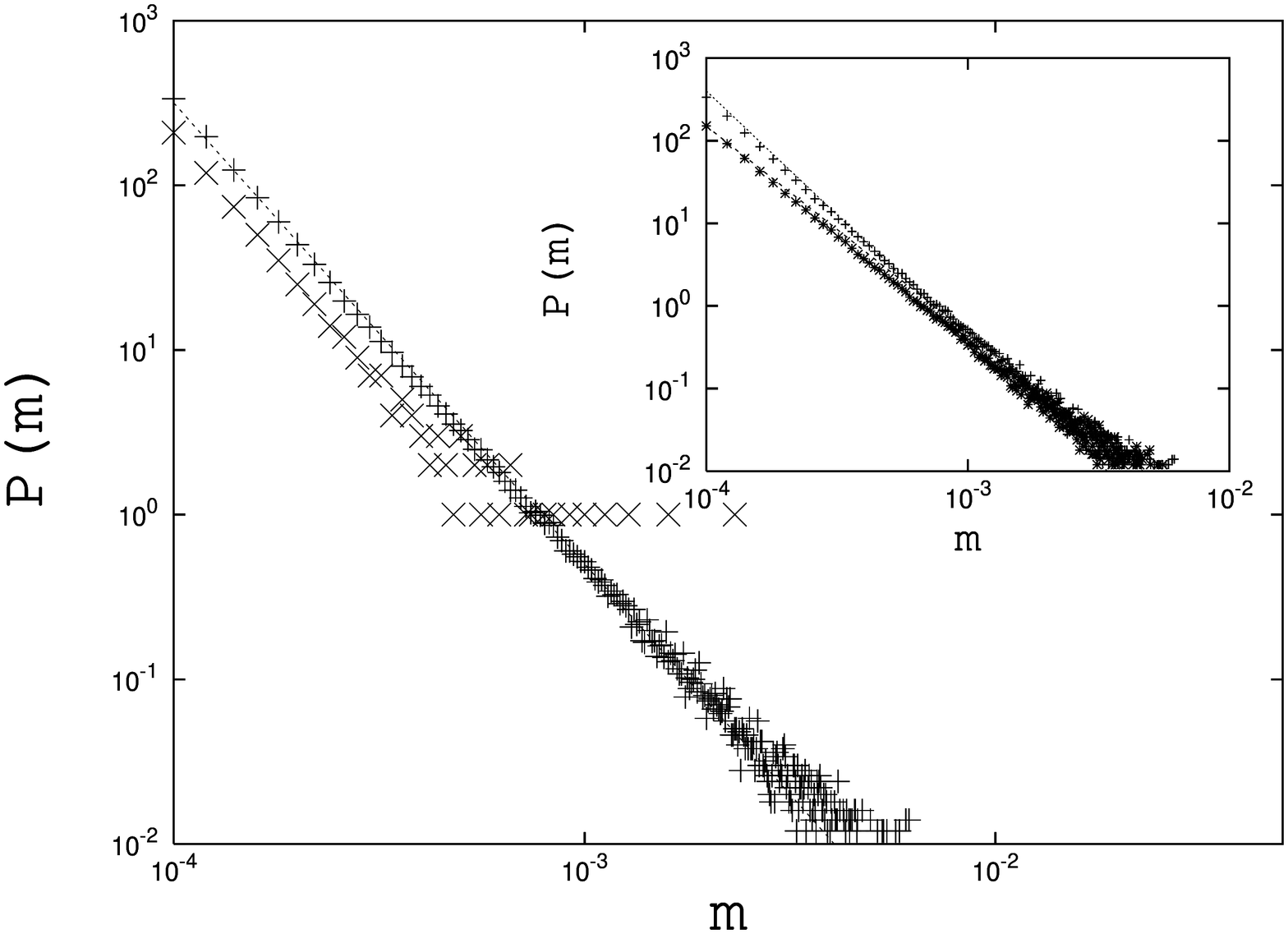}}} \par}
\vspace{0.3cm}

\noindent \textbf{\footnotesize Fig. 3}{\footnotesize : The log-log
plot of the average avalanche size distributions \( P(m) \) against
\( m \) for \( N=50,000 \) with \( d\sigma =1/N \) for strictly
uniform fiber strength distribution (cross) and uniform-on-average
fiber strength distribution (averaged over 501 bundle realisations;
plus). The dotted line has a slope \( \eta =-3.0 \), representing
eqn. (17). The inset shows the avalanche size distributions for uniform-on-average
fiber strength distribution, when (a) the external load increases
by a fixed amount \( d\sigma =1/N \) (plus) and (b) the avalanches
are triggered by breaking the next weakest fiber (star).The dotted
and dashed lines in the inset correspond to \( \eta =-3.0 \) and
\( \eta =-2.5 \) respectively. }{\footnotesize \par}

\vskip .05in

This result (17) for the avalanche size distribution \( P(m) \) is
therefore valid for other distributions of fiber strength (cf. \cite{Pach2000}),
when the avalanche size is defined through (15). If one looks for
the statistics of avalanches initiated by breaking the next weakest
fiber in bundles with uniform-on-average fiber strength distribution,
as in Hemmer et al \cite{HH92,D92}, then one gets \( \eta =5/2 \).
This is shown in the inset of Fig. 3, where the avalanches are defined
in both ways: with fixed increase in \( \sigma  \) (giving \( \eta =3.0 \))
and by breaking the next weakest fiber (giving \( \eta =2.5 \)).
The difference in the above exponent values therefore originates from
different ways of defining the avalanches; in our method of defining
avalanches here, the external load on the bundle increases uniformly,
while in the other method the external load increase has intrinsic
fluctuations due to the randomness of the fiber strengths and the
restriction on initiating the avalanches by breaking only the next
weakest fiber.

\vskip.1in

\noindent \textbf{VII. Plastic deformation and stress-strain relation}

\noindent One can now consider a slightly modified strength distribution
of such a fiber bundle, showing plastic-deformation characteristics
\cite{Dan45,More94}. For this, we consider a RFB strength distribution,
having a lower cutoff. Until failure of any of the fibers (due to
this lower cutoff), the bundle shows linear elastic behavior. As soon
as the fibers start failing, the stress-strain relationship becomes
nonlinear. The dynamic critical behavior remains essentially the same
and the static (fixed point) behavior shows elastic-plastic deformation
before rupture of the bundle. 

Here the fibers are elastic in nature having identical force constant
\( \kappa  \) (see Fig. 1) and the random fiber strengths distributed
uniformly in the interval \( [\sigma _{L},1] \) with \( \sigma _{L}>0 \);
the normalised distribution of the threshold stress of the fibers
thus has the form (see Fig. 4): \begin{equation}
\label{jan31}
\rho (\sigma )=\left\{ \begin{array}{cc}
0, & 0\leq \sigma \leq \sigma _{L}\\
\frac{1}{1-\sigma _{L}}, & \sigma _{L}<\sigma \leq 1
\end{array}\right\} .
\end{equation}

\vspace{0.3cm}
{\centering \resizebox*{8cm}{6cm}{\includegraphics{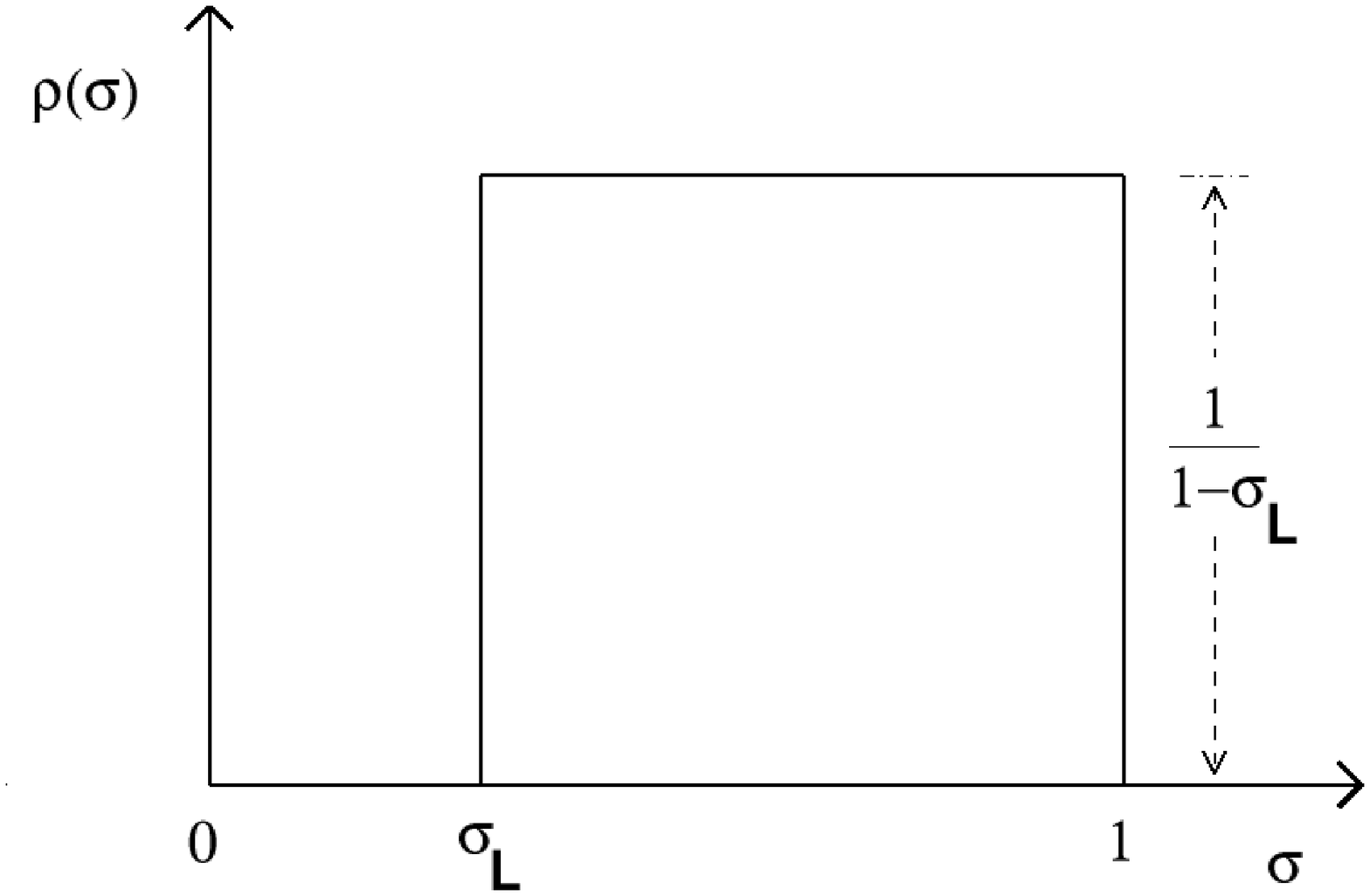}} \par}
\vspace{0.3cm}

\noindent \textbf{\footnotesize Fig.} {\footnotesize 4: The fiber
breaking strength distribution \( \rho (\sigma ) \) considered for
studying elastic-plastic deformation behavior of the RFB model. This
distribution gives the recurrence relation (19).}{\footnotesize \par}

\vskip .1in

For an applied stress \( \sigma \leq \sigma _{L} \) none of the fibers
break, though they are elongated by an amount \( \delta  \) \( =\sigma /\kappa  \).
The dynamics of breaking starts when applied stress \( \sigma  \)
becomes greater than \( \sigma _{L} \). Now, for \( \sigma >\sigma _{L} \)
the fraction of unbroken fibers follows a recurrence relation (for
\( \rho (\sigma ) \) as in Fig. 4): \begin{equation}
\label{abcdef}
U_{t+1}(\sigma )=1-\left[ \frac{F}{NU_{t}(\sigma )}-\sigma _{L}\right] \frac{1}{1-\sigma _{L}}=\frac{1}{1-\sigma _{L}}\left[ 1-\frac{\sigma }{U_{t}(\sigma )}\right] ,
\end{equation}

\noindent which has stable fixed points: \begin{equation}
\label{feb7c}
U^{\star }(\sigma )=\frac{1}{2(1-\sigma _{L})}\left[ 1+\left( 1-\frac{\sigma }{\sigma _{c}}\right) ^{1/2}\right] ;\sigma _{c}=\frac{1}{4(1-\sigma _{L})}.
\end{equation}

\noindent The RFB model now has a critical point \( \sigma _{c}=1/[4(1-\sigma _{L})] \)
beyond which total failure of the bundle takes place. The above equation
also requires that \( \sigma _{L}\leq 1/2 \) (to keep the fraction
\( U^{\star }\leq 1 \)). As one can easily see, the dynamics of \( U_{t}(\sigma ) \)
for \( \sigma <\sigma _{c} \) and also at \( \sigma =\sigma _{c} \)
remains the same as discussed in the earlier section. At each fixed
point there will be an equilibrium elongation \( \delta (\sigma ) \)
and a corresponding stress \( S=U^{\star }\kappa \delta (\sigma ) \)
develops in the system (bundle). This \( \delta (\sigma ) \) can
be easily expressed in terms of \( U^{\star }(\sigma ) \). This requires
the evaluation of \( \sigma ^{\star } \), the internal stress per
fiber developed at the fixed point, corresponding to the initial (external)
stress \( \sigma  \) (\( =F/N \)) per fiber applied on the bundle
when all the fibers were intact. From the first part of eqn. (19),
one then gets (for \( \sigma >\sigma _{L} \))\begin{equation}
\label{feb7}
U^{\star }(\sigma )=1-\frac{\sigma ^{\star }-\sigma _{L}}{(1-\sigma _{L})}=\frac{1-\sigma ^{\star }}{1-\sigma _{L}}.
\end{equation}
 Consequently, \begin{equation}
\label{feb7}
\kappa \delta (\sigma )=\sigma ^{\star }=1-U^{\star }(1-\sigma _{L}).
\end{equation}
 It may be noted that the internal stress \( \sigma _{c}^{\star } \)
is universally equal to \( 1/2 \) (independent of \( \sigma _{L} \))
at the failure point \( \sigma =\sigma _{c} \) of the bundle. This
finally gives the stress-strain relation for the RFB model : \begin{equation}
\label{jan31}
S=\left\{ \begin{array}{cc}
\kappa \delta , & 0\leq \sigma \leq \sigma _{L}\\
\kappa \delta (1-\kappa \delta )/(1-\sigma _{L}), & \sigma _{L}\leq \sigma \leq \sigma _{c}\\
0, & \sigma >\sigma _{c}
\end{array}\right\} .
\end{equation}

\vspace{0.3cm}
{\centering \resizebox*{11cm}{8cm}{\includegraphics{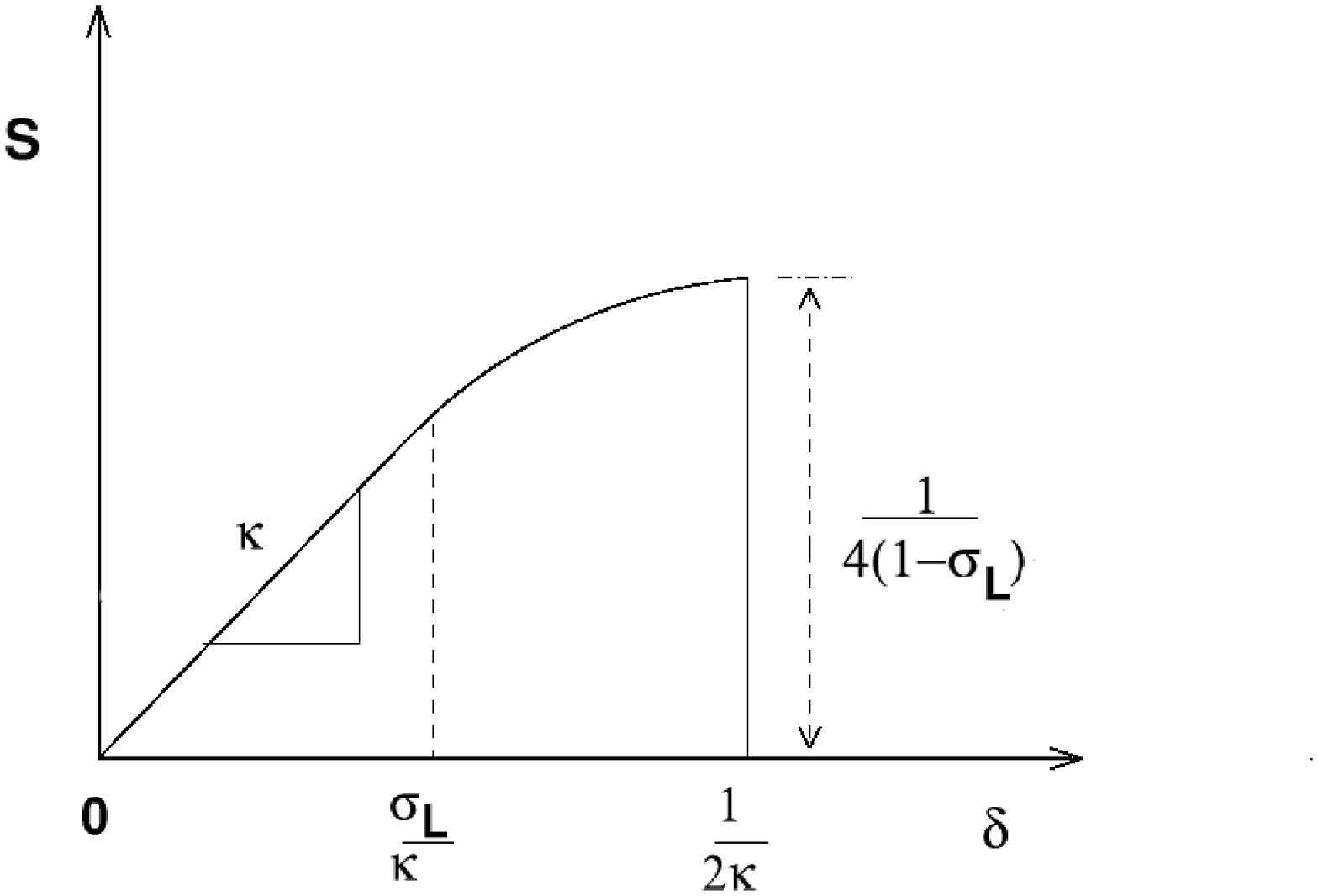}} \par}
\vspace{0.3cm}

\noindent \textbf{\footnotesize Fig. 5:} {\footnotesize Schematic
stress (\( S \))-strain (\( \delta  \)) curve of the bundle (shown
by the solid line), following eqn. (23), with the fiber strength distribution
(18) (as shown in Fig. 4). }{\footnotesize \par}

\vskip .1in

\noindent This stress-strain relation is schematically shown in Fig.
5, where the initial linear region has slope \( \kappa  \) (the force
constant of each fiber). This Hooke's region for stress \( S \) continues
up to the strain value \( \delta =\sigma _{L}/\kappa  \), until which
no fibers break (\( U^{\star }(\sigma )=1 \)). After this, nonlinearity
appears due to the failure of a few of the fibers and the consequent
decrease of \( U^{\star }(\sigma ) \) (from unity). It finally drops
to zero discontinuously by an amount \( \sigma _{c}^{\star }U^{\star }(\sigma _{c})=1/[4(1-\sigma _{L})]=\sigma _{c} \)
at the breaking point \( \sigma =\sigma _{c} \) or \( \delta =\sigma ^{\star }_{c}/\kappa =1/2\kappa  \)
for the bundle. This indicates that the stress drop at the final failure
point of the bundle is related to the extent (\( \sigma _{L} \))
of the linear region of the stress-strain curve of the same bundle. 

Here, the plasticity (nonlinearity) in the response of the bundle
comes naturally from partial failure of the fibers (and the consequent
redistribution of stress among the surviving fibers), after the assumed
linear region until the lower threshold \( \sigma _{L} \) of failure
(18). The total failure of the bundle is again discontinuous here
and the entire nonlinear response characteristics is analytically
calculable in this simple model.

\noindent \newpage

\noindent \textbf{VIII. Summary and conclusions}

\noindent We have reported here an analytic study of the failure dynamics
and the consequent plastic deformation characteristics of the random
fiber bundle model having the property of global load sharing. This
has been done here for uniform strength distribution \( \rho (\sigma ) \)
of fibers in the bundle (up to a cutoff). As mentioned before, this
has been possible due to the inherent mean-field nature of the model.
The recurrence relation (1) captures essentially all the intriguing
features of the dynamics. We found that both the breakdown susceptibility
\( \chi  \) and the relaxation time \( \tau  \) diverge as the applied
stress \( \sigma  \) approaches its global failure point \( \sigma _{c} \)
(\( =1/4 \) for the uniform strength distribution as shown in Fig.
2) from below, with the same exponent value \( \gamma =\alpha =1/2 \).
The critical dynamics of the model follows a strict power law decay
at \( \sigma =\sigma _{c} \): \( U_{t}-1/2\propto t^{-1} \). Though
we have identified \( O\equiv U^{\star }(\sigma )-1/2 \) as the order
parameter (with exponent \( \beta =1/2 \)) for the continuous transition
in the model, unlike conventional phase transitions it does not have
a real-valued existence for \( \sigma >\sigma _{c} \). From finite-size
scaling study, we see that there is a correlation length which diverges
with an exponent \( \nu =1 \), as \( \sigma _{c} \) is approached
from below. The avalanche size distribution \( P(m) \) for this mean-field
dynamics of the RFB model is given by \( P(m)\sim m^{-\eta } \),
\( \eta =3 \). This has been confirmed here numerically.  As mentioned
before, this result is valid for the avalanche sizes defined through
(16), where the external load on the bundle increases uniformly until
the total failure at \( \sigma _{c} \). The present as well as the
earlier numerical results \cite{PS97,Pach2000,SB01} all confirm that
the analytic results for the exponents \( \alpha  \), \( \gamma  \)
and \( \eta  \) (for \( \tau ,\chi  \) and \( P(m) \) respectively)
are not necessarily restricted to the uniform distribution of fiber
strength (assumed here) and are more generally valid.  The model also
shows realistic plastic deformation behavior with a shifted (by \( \sigma _{L} \),
away from the origin) uniform distribution of fiber strengths. The
stress-strain curve for the model clearly shows three different regions:
elastic or linear part (Hooke's region) when none of the fibers break
(\( U^{\star }(\sigma )=1 \)), plastic or nonlinear part due to the
successive failure of the fibers (\( U^{\star }(\sigma )<1 \)) and
then finally the stress drops suddenly (due to the discontinuous drop
in the fraction of surviving fibers from \( U^{\star }(\sigma _{c}) \)
to zero) at the bundle failure point at \( \sigma =\sigma _{c} \)
(\( =1/[4(1-\sigma _{L})] \) for the failure strength distribution
(18)). Simplicity of the model and consequently of the recurrence
relation for the breaking dynamics allows it to have exact analytic
results for all its static and dynamic behaviors of breaking and the
resulting plasticity.

\newpage
\end{document}